\documentclass[ reprint,superscriptaddress, amsmath,amssymb, aps, twocolumn,pra, nofootinbib]{revtex4-1}
\pdfoutput=1
\usepackage[dvipsnames]{xcolor}

\usepackage[normalem]{ulem}
\usepackage{multirow}
\usepackage{enumitem}

\usepackage{graphicx}
\usepackage{dcolumn}
\usepackage{bm}
\usepackage{hyperref}
\hypersetup{colorlinks = true, linkcolor = [rgb]{0.19411,0.51882,0.667058}, urlcolor = [rgb]{0.125490,0.29542,0.1647058}, citecolor = [rgb]{0.75882,0.37411,0.14117}}
\usepackage{makecell}

\usepackage{setspace}
\usepackage{eurosym}
\usepackage{amssymb}

\usepackage{braket}
\usepackage[T1]{fontenc}

\graphicspath{{../images/}}
\usepackage[toc,page]{appendix}

\newcommand{\nn}{\nonumber \\}

\usepackage{braket}
\usepackage[bottom]{footmisc}
\def\>{\rangle}
\def\<{\langle}

\usepackage{mathtools}
\DeclarePairedDelimiter{\ceil}{\lceil}{\rceil}

\usepackage[bottom]{footmisc}
\begin{document}

\title{
Experimental implementation of secure anonymous protocols on an eight-user quantum network
}

\author{Zixin Huang$^\ddag$}
\email{zixin.huang@sheffield.ac.uk}
\affiliation{Department of Physics and Astronomy, The University of Sheffield}
\thanks{$^\ddag$These authors contributed equally to this work}

\author{Siddarth Koduru Joshi$^\ddag$}
\email{SK.Joshi@Bristol.ac.uk}
\affiliation{Quantum Engineering Technology Labs, H. H. Wills Physics Laboratory \& Department of Electrical and Electronic Engineering, University of Bristol}

\author{Djeylan Aktas}
\affiliation{Quantum Engineering Technology Labs, H. H. Wills Physics Laboratory \& Department of Electrical and Electronic Engineering, University of Bristol}

\author{Cosmo Lupo}
\affiliation{Department of Physics and Astronomy, The University of Sheffield}
\author{Armanda O. Quintavalle}
\affiliation{Department of Physics and Astronomy, The University of Sheffield}

\author{Natarajan Venkatachalam}
\affiliation{Quantum Engineering Technology Labs, H. H. Wills Physics Laboratory \& Department of Electrical and Electronic Engineering, University of Bristol}

\author{S\"oren Wengerowsky}
 \affiliation{Institute for Quantum Optics and Quantum Information - Vienna (IQOQI) \& Vienna Center for Quantum Science and Technology (VCQ), Vienna, Austria}
\affiliation{Currently at ICFO-Institut de Ciencies Fotoniques, The Barcelona Institute of Science and Technology, 08860 Castelldefels  (Barcelona), Spain}

\author{Martin Lon\v{c}ari\'{c}}
\affiliation{Photonics and Quantum Optics Research Unit, Center of Excellence for Advanced Materials and Sensing Devices, Ru\dj{}er Bo\v{s}kovi\'{c} Institute, Zagreb, Croatia}
\author{Sebastian Philipp Neumann}
\affiliation{Institute for Quantum Optics and Quantum Information - Vienna (IQOQI) \& Vienna Center for Quantum Science and Technology (VCQ), Vienna, Austria}
\author{Bo Liu}
\affiliation{College of Advanced Interdisciplinary Studies, NUDT, Changsha, 410073, China}
\author{\v{Z}eljko Samec}
\affiliation{Photonics and Quantum Optics Research Unit, Center of Excellence for Advanced Materials and Sensing Devices, Ru\dj{}er Bo\v{s}kovi\'{c} Institute, Zagreb, Croatia}
\author{Laurent Kling}
\affiliation{Quantum Engineering Technology Labs, H. H. Wills Physics Laboratory \& Department of Electrical and Electronic Engineering, University of Bristol}
\author{Mario Stip\v{c}evi\'{c}}
\affiliation{Photonics and Quantum Optics Research Unit, Center of Excellence for Advanced Materials and Sensing Devices, Ru\dj{}er Bo\v{s}kovi\'{c} Institute, Zagreb, Croatia}
\author{Rupert Ursin}
\affiliation{Institute for Quantum Optics and Quantum Information - Vienna (IQOQI) \& Vienna Center for Quantum Science and Technology (VCQ), Vienna, Austria}

\author{John G. Rarity}
\affiliation{Quantum Engineering Technology Labs, H. H. Wills Physics Laboratory \& Department of Electrical and Electronic Engineering, University of Bristol}

\begin{abstract}

Anonymity in networked communication is vital for many
privacy-preserving tasks. Secure key distribution alone is insufficient for high-security communications. Often, knowing who transmits a message to whom and when must also be kept hidden from an adversary. 
Here we experimentally demonstrate 5 information-theoretically secure anonymity protocols on an 8 user city-wide quantum network using polarisation-entangled photon pairs. At the heart of these protocols is anonymous broadcasting, which is a cryptographic primitive that allows one user to reveal one bit of information while keeping her identity anonymous.  For a network of $n$ users, the protocols retain anonymity for the sender, given less than $n-2$ users are dishonest. 
This is one of the earliest implementations of genuine multi-user cryptographic protocols beyond standard QKD. Our anonymous protocols enhance the functionality of any fully-connected Quantum Key Distribution network without trusted nodes.
 
\end{abstract}

\date{\today}

\maketitle

\section{introduction}
Quantum cryptography is one of the fastest-growing quantum technologies. Quantum Key Distribution (QKD) has been demonstrated across a huge spectrum of platforms \cite{scarani2009security,pirandola2019advances}, and proof-of-principle demonstrations are quickly being adapted into commercial prototypes. Recent experimental progress and the development of potentially large-scale networks \cite{bradley2019ten,yin2017satellite,sun2016quantum,Soeren2018,joshi2020trusted} open up the possibility of a full-scale quantum internet \cite{wehner2018quantum,kimble2008quantum,li2011information}.
Many protocols have been developed for multi-user quantum networks, such as secret voting \cite{PhysRevA.75.012333,hillery2006towards}, secret sharing \cite{xiao2004efficient,hillery1999quantum}, clock synchronization \cite{komar2014quantum}, and distributed blind quantum computation \cite{beals2013efficient}; these all using multipartite states. However, multipartite states are very complex to create and the performance of protocols based on such states can degrade rapidly with losses (i.e., distance). Protocols based on bipartite states, such as the ones presented here, are often the simplest and best choice given any realistic amount of loss. Further, they are compatible with today's state-of-the-art quantum networks which distribute bipartite entanglement and use this to perform quantum communication.

As we build the world's most secure networks, it is important to ensure that traffic on these networks is anonymous in addition to being impossible to decrypt. 
Thus, anonymity as a cryptographic primitive is becoming increasingly important for networked applications due to concerns for privacy and censorship. Applications include   anonymous communication, secret auctions \cite{stajano1999cocaine}, and
anonymous crypto-currency transactions \cite{PhysRevA.97.032345}.
The first anonymous broadcating protocol was named the ``cryptographer's dining problem'' \cite{chaum1988dining}. There, $n$ users establish shared secret keys with all the other participants, allowing one participant to send a single bit of classical information whilst keeping her identity secret. 

A traceless and efficient quantum anonymous broadcasting was reported by Christandl and Wehner \cite{roy2005advances}. 
There, a trusted resource distributes ahead
of time an $n$-partite entangled GHZ state
\mbox{$\ket\psi_\mathrm{GHZ} = \frac{1}{\sqrt{2}} \left( \ket{0}^{\otimes n} + \ket{1}^{\otimes n} \right)$}. 
Later, Broadbent et al.~derived a class of information-theoretically secure protocols that allow one party to transmit a message anonymously without an honest majority \cite{broadbent2007information}. 
The key enabler for the anonymity protocols in Ref.~\cite{broadbent2007information} is anonymous broadcasting (also known as parity), which requires authenticated pairwise private channels.

While GHZ state-based  (or in general, multipartite state-based) networks are difficult to scale-up to many users, our approach exploits a scalable, fully-connected, metropolitan-sized quantum network test bed~\cite{joshi2020trusted}.
In our setup, we use this quantum network to distribute pair-wise secret keys between all the participants. 

After obtaining these keys, we solve the classical ``cryptographer's dining problem'', and implement the following, initially proposed in \cite{broadbent2007information}.
\begin{enumerate}
\item \textit{Anonymous broadcasting (parity)}, allows a single user to transmit one bit of information.
\item \textit{Veto}, allows a single participant to unilaterally stop or pass a binary decision-making task whilst remaining anonymous.
\item \textit{Notification}, here a participant notifies a list of others, but revealing neither the number of participants nor their identity.
\item \textit{Collision detection}, which verifies whether there is a single sender.
\item \textit{Anonymous private message transmission}, allowing a sender anonymously transmit a message to a receiver, despite possible malicious interference from other users.
\end{enumerate}

Anonymous broadcasting serves as a fundamental building block for the other protocols implemented here, as well as many network anonymous protocols  \cite{broadbent2007information}.
In the classical anonymous broadcasting protocol, each participant shares a secret key bit with every other participant. Each participant outputs the modulo-2 sum (parity) of all the key bits she shares. If the sender wants to output the bit 0, she does nothing, and if she wants to output the bit 1, she inverts the output. Assuming only one user wants to communicate, and since all the keys enter the sum twice, the net parity should be 0 unless one bit was inverted.

The structure of the paper follows. In Sec.~\ref{sec:exp} we will describe the quantum network we used to distribute unconditionally secure secret keys, then in Sec.~\ref{sec:protocols} we describe in detail the anonymous multi-user network protocols we executed using the quantum network. We present our results in Sec.~\ref{sec:results}.

\section{The quantum network}\label{sec:exp}
\begin{center}
\begin{figure}
\includegraphics[width=0.95\columnwidth]{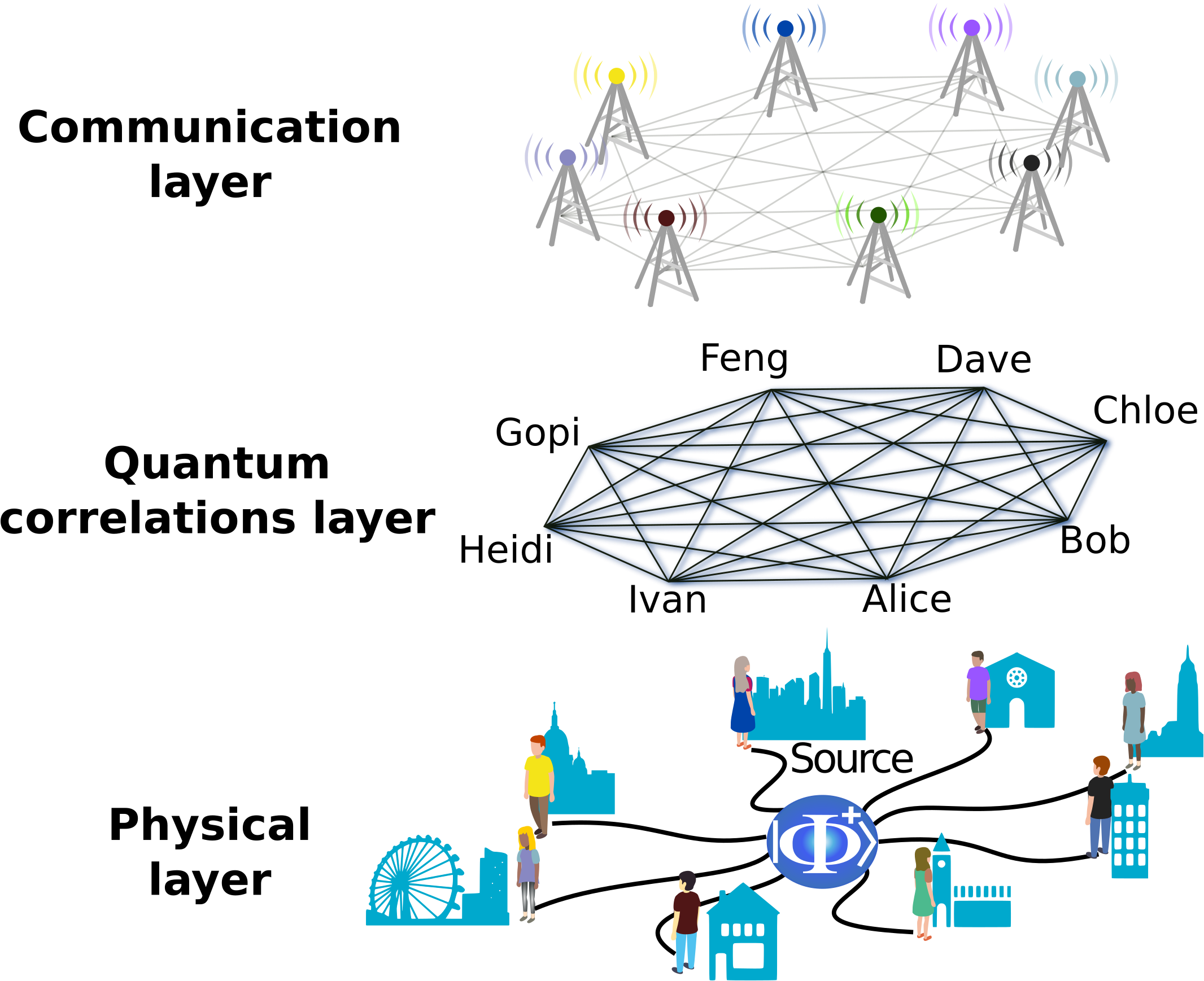}
\caption{\textit{Overview of the quantum network test bed}. The physical layer consists of a broadband source of bipartite polarisation entanglement that is multiplexed and distributed to several users via one fibre each. The users make measurements of the photons they receive and share entanglement between all users in the quantum correlation layer. Finally, in the communication layer, they use the protocols described in this paper to anonymously broadcast,  veto, notify a chosen user, or transmit a message. } \label{fig:layers}  
\end{figure}
\end{center}

The network architecture can be divided into different layers of abstraction. The hardware is part of the physical layer. The processing of measurements to realise entanglement and quantum secure keys is part of the ``quantum correlations layer''.
All our anonymous protocols were run within the communication layer of a quantum network (see  Fig.~\ref{fig:layers}). The experiment was implemented using a
 8-node city-scale quantum network based on bipartite entanglement distribution described in Ref.~\cite{joshi2020trusted}. Each of the 8 users shares a different bipartite entangled state with every other user, forming the fully connected graph topology of the quantum network our protocols require. Further, since every node directly shares entanglement with every other node, we do not make use of trusted nodes. 
 
\begin{center}
\begin{figure}
\includegraphics[width=0.95\columnwidth]{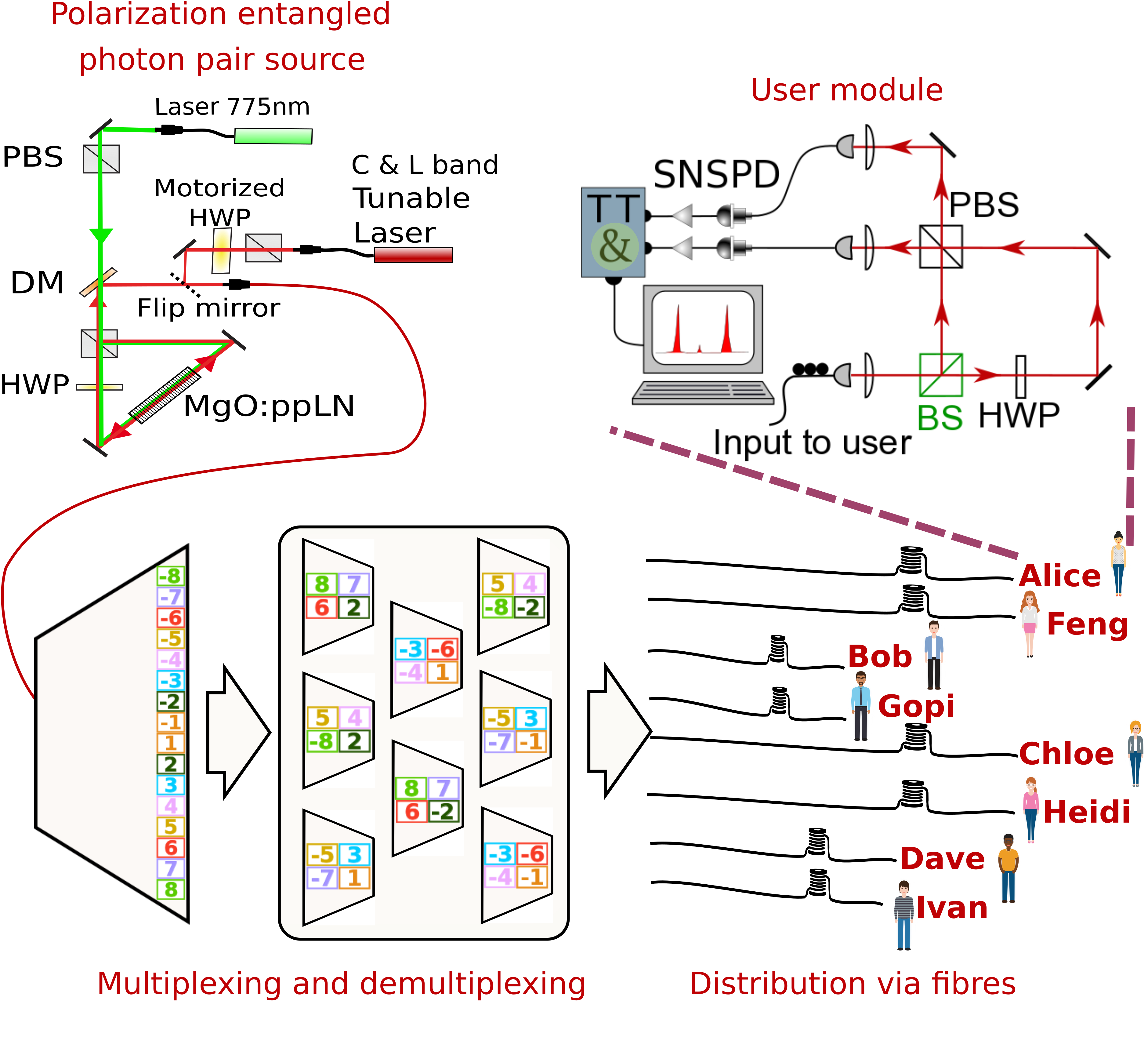}
\caption{\textit{Experimental setup}. The entangled photon pair source is based on a type-0 MgO:ppLN crystal pumped bidirectionally by a 775.1~nm laser. A flip-mirror and motorised half wave plate (HWP) are used to couple a tunable telecom laser into the output fibre for polarisation control. The multiplexers/demultiplexers combine the shown combination of wavelength channels and distribute them to all 8 users. The channel numbers are shown here plus 34 correspond to the standard ITU 100\,GHz DWDM grid. Due to energy conservation photons in channels \{1,-1\}, \{2,-2\}, and so on are the only ones entangled. Each user is connected to the central hub by one fibre. The figure also shows the setup of each user's polarisation analysis and detection module.  (PBS: Polarisation Beam Splitter, DM: Dichroic Mirror, SNSPD: Superconducting Nanowire Single Photon Detector, TT: Time Tagger, BS: non-polarising Beam Splitter. } \label{fig:setup}  
\end{figure}
\end{center}

 The physical layer is shown in more detail in Fig.~\ref{fig:setup}. It consists of the entanglement source, the multiplexers and demultiplexers needed to distribute the entanglement, a single transmission fibre for each user, and each user's detection module with the single-photon detectors. 
 The entangled states are all produced by a single broadband source of polarization-entangled photon pairs. Polarisation entanglement is used to simplify each user's measurement module and is also compatible with long distance passively stable distribution over deployed fibres~\cite{wengerowsky2019entanglement,wengerowsky2020passively}. The output is de-multiplexed into 16 wavelength channels using International Telecommunications Union (ITU) 100 GHz top hat Dense Wavelength Division Multiplexing (DWDM) channels. The source itself is based on spontaneous parametric down-conversion of a 775.1\,nm pump beam in a Magnesium oxide doped Periodically Poled Lithium Niobate (MgO:ppLN) crystal within a Sagnac interferometer, producing degenerate entangled photons centred at 1550.2\,nm (which coincides with the 100\,GHz (i.e.,0,8\,nm) wide ITU channel Ch34 at 1550.12\,nm) with a Full Width at Half Maximum bandwidth of 
 $\approx 60$\,nm. Due to energy conservation during down-conversion, pairs are only found in channels at equal spectral distance from the central frequency. A 50-50 fibre beam splitter sends photons randomly to one of two output ports (which can be thought of as passive time-division multiplexing). A multiplexing stage combines the channels together such that every user receives four frequency channels containing photons whose entangled partner photons were routed to four other users. 
 
 Each user measures all the photons they receive in either the horizontal/vertical (HV) or the diagonal/anti-diagonal (DA) polarisation basis. The modules to perform this measurement are shown in Fig~\ref{fig:setup}. Note that each user performs a passive basis choice using a 50:50 beam splitter and utilises just two detectors (rather than the typical 4) by introducing a delay of 3.7\,ns for detection events in the DA basis. 
 
 In the quantum correlation layer (see Fig.~\ref{fig:layers}), the above measurement outcomes are compared to generate a secret key following the BBM92 protocol~\cite{BBM92} of quantum key distribution.
To generate the secret keys, the users exchanged arrival time information. Since our network uses a unique 2-detector measurement scheme (see Ref~\cite{joshi2020trusted} for implementation and security proof) the basis choice reconciliation information can be extracted from the temporal cross-correlation of the arrival time of photon detection events at any two users. These sifted keys are error corrected using a Low-Density Parity Check (LDPC) error correction code. Every user runs multiple instances of the key generation program to generate keys from the sifted data following the BBM92 protocol. We used a security parameter of $10^{-5}$ for this implementation. Users maintain a separate key store for every other user. 

Lastly, our protocols are executed in the communication layer. To perform any of our anonymous protocols, users consume the desired number of bits from the appropriate key stores. Keys from key stores are used only once.

\section{Anonymous protocols}
\label{sec:protocols}

Having established secure keys, the users can now perform multiple-party network protocols. We perform 5 protocols altogether, and in this section, we describe three in detail: anonymous broadcasting, veto, and anonymous private message transmission. The notification and collision detection protocols are described in the appendix. An advantage of the veto and notification protocol is that they do not suffer from collision problems while anonymous broadcasting does.

\subsection{Anonymous broadcasting (parity)}

Assuming only one user wants to communicate, the anonymous broadcasting protocol allows one out of $n$ participants to broadcast one bit at a time. 
If the speaker wishes to communicate 0, she does nothing; if she wishes to communicate a 1, then she inverts \textit{one} bit of the key she shares with another participant, out of $(n-1)$. She does this without informing any other participant. Details follow:

\begin{enumerate}
    \item Using a secure bit in the string extracted from the QKD sub-protocol, the user who wants to make an announcement flips one of the bits (out of $(n-1)$). 
    \item Each participant then adds and announces the modulo-2 sum (parity) of her $(n-1)$ secret bits.
    \item The overall parity sum of every participant's broadcast bit value is then computed. Since every secret bit enters the addition process exactly twice if no one transmits or the transmitted bit is 0, the parity of the outputs is 0, and if one participant transmits the bit 1, the parity is 1. 
\end{enumerate}

Here the purpose of the quantum states is to provide the unconditionally secure secret keys between each of the participants, and the rest is classical. Ref.~\cite{chaum1988dining} provides an information-theoretic proof of the security of the anonymous broadcasting protocol, given that each participant share a private communication channel, i.e.\ a secure key with one another. Each round of the anonymous broadcasting protocol consumes in total $n(n-1)/2$ secret bits.

\textit{Anonymity:} Intuitively, the protocol is anonymous for the following reasons: since a bit-flip is applied locally, no transmissions are necessary to change the overall parity. The only information made public is the overall parity which is correlated with all other $(n-1)$ users in the graph.  

\textit{Security:}  
If all $(n-1)$ participants fully cooperate against one, then no protocol can keep the sender's identity secret. However, if there are $t$ colluding users, then all other participants are equally likely to be the sender. The bit flip is hidden by a one-time pad -- therefore the sender is protected.


So far, we have assumed that only a single user is sending the bit $1$ in any round. If more than one participant inverts their bit, this will lead to a collision error. 
We now calculate the probability of error due to collision. We assume that, given that one participant has sent the bit 1, there is a probability $p$ that every other users also wants to communicate the bit 1. 
Then the probability of error $P$ in the final parity is dictated by having an odd number of additional speakers. 
The probability that $i$ users want to communicate and $n-i-1$ do not is $p^i (1-p)^{n-i-1}$. Summing over odd values of $i$ and over all permutations of the users, we obtain
\begin{align}
P &= \sum_{i \text{ odd} }^{n-1}  {n-1 \choose i}p^i (1-p)^{n-1-i} \nn
    &=\frac{1}{2}-\frac{1}{2}(1-2p)^{n-1}.
\end{align}

\noindent This issue due to collision becomes more problematic as the number of users become large. Our collision detection protocol in Appendix~\ref{sec:collision} addresses this issue.

Armed with the anonymous broadcasting protocol, we now proceed to implement a number of other anonymous protocols. 
Note that the \textit{veto} and \textit{notification} protocols serve as the building blocks for \textit{collision detection} and \textit{anonymous private message transmission} \cite{broadbent2007information, broadbent2010exact}.

\subsection{Anonymous Veto}

In the veto protocol, each of the $n$ participants decides on a binary-decision problem by submitting a classical bit \mbox{$x_i \in \{0,1\}$}.  If the $i^{\mathrm{th}}$ participant chooses to veto, then $x_i =1$, otherwise $x_i =0$.
The final output is $1$ if at least one participant submits the bit $1$: the output of the veto protocol is effectively the logical OR function of all of the 
$x_i$'s, allowing all participants to cast a vote. 

The veto protocol is neither vulnerable to collisions, nor to malicious jamming. 
In principle, the anonymous broadcasting protocol can also be used for a decision-making task.  However, it is vulnerable to collisions and malicious jamming.
In the event of a collision with an even number of participants transmitting the bit 1, the parity of the overall output will be 0 (non-veto).
If the broadcasting channel is not simultaneous, the last participant to announce their parity can always choose their input such that the overall parity is 0 (non-veto), which constitutes malicious jamming.

\noindent 
\begin{figure}[t]
\includegraphics[width=0.9\columnwidth]{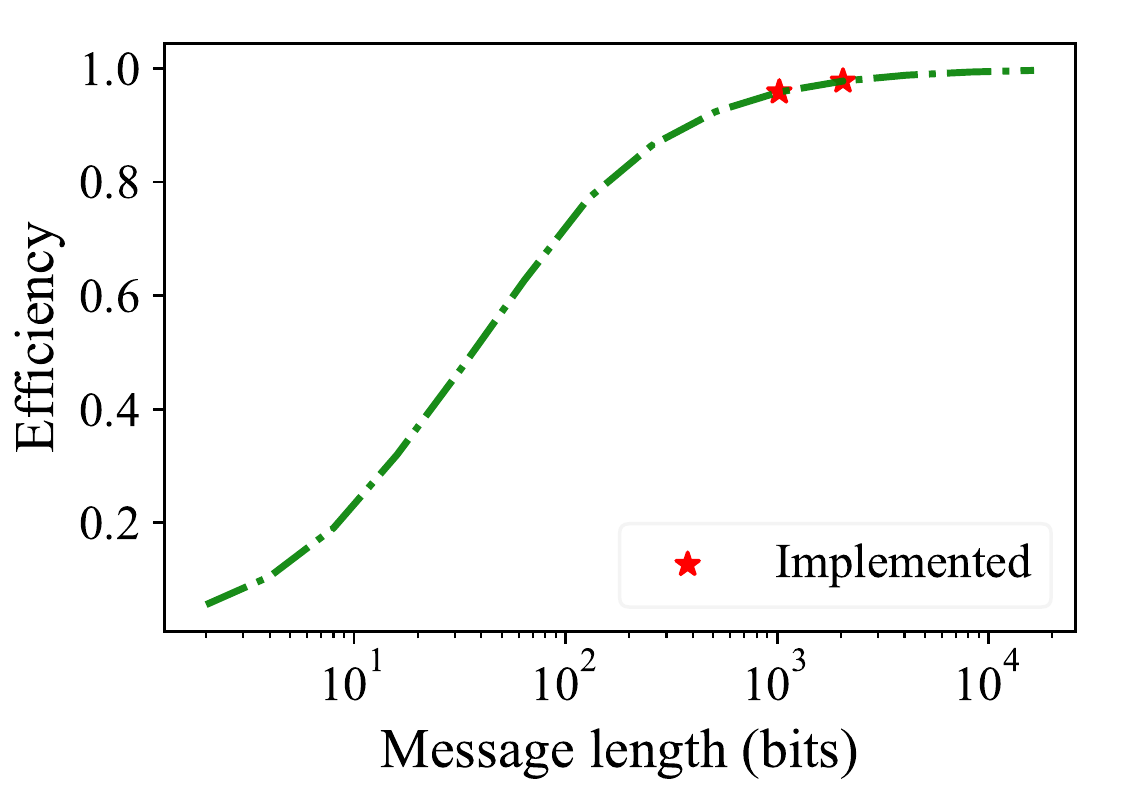}
\caption{\textit{Efficiency of the encoding} for anonymous private message transmission, $m/(m+2\gamma)$ for different message lengths, given a fixed security parameter $\beta = 16$.} 
\label{fig:eff}    
\end{figure}

Both these problems are solved by the veto protocol.
For the veto protocol to be valid, the participants have to agree that if one participant vetoes, all participants accept the veto outcome, and if the outcome is a veto, all participants accept that someone
has vetoed \cite{hao2009power}. 
%
%
We define $\beta$ to be a user defined parameter chosen and agreed upon by all participants to ensure the desired success probability of each protocol (see Table~\ref{tab:sum}).
The protocol follows. 
\begin{enumerate}
    \item The $n$ participants agree on $n$ orderings such that each ordering has a different last participant. This step ensures that every participant has a chance to be the last to broadcast. 
    \item For each ordering, the following is repeated $\beta$ times:
    \begin{enumerate}
        \item Depending on $x_i$, each participant sets the value of another bit, $c_i$, in the following way: if $x_i = 0$, then $c_i$ = 0; 
        otherwise if $x_i = 1$, then $c_i$ is chosen randomly.
        \item The participants execute the anonymous broadcasting protocol with inputs 
        $\{c_i\}$. That is, if $c_i = 0$, they do nothing, and if $c_i = 1$, they flip the bit value of one of their modulo-2 sum.
        \item  If the parity at any round is 1, or if any participant refuses to broadcast, then the veto result is set to 1 (veto).
    \end{enumerate}
\end{enumerate}

\color{black}
If for any of the $\beta$ rounds the output is 1, then we know someone has vetoed. The value of $c_i$ is randomly chosen to be either 0 or 1, so that if an even number of people want to veto, they do not end up with an even number of collisions all the time.

A single round of the veto protocol consumes at most $\beta n^2(n-1)/2$ secret bits and succeeds with probability at least $1-2^{-\beta}$.
If all participants in the protocol have $x_i =0$, then the inputs to the anonymous broadcasting protocols are $c_i = 0$, then the output of the protocol is 0 with probability 1. 
If there is one participant with $x_i =1$, then let's suppose that the sender speaks last. If the goal of an adversary is to set the output to be 0, then it is unfruitful for them to perform any action preemptively. 
Now, because of the way that $c_i$ is chosen, then with probability at least $1-2^{-\beta}$ (if everyone is honest, the probability of success is $1-2^{-n\beta}$), the output of the protocol will be 1. The identity of the sender is hidden by the anonymous broadcasting protocol.

\subsection{Anonymous private message transmission}

The anonymous private message transmission allows a sender to anonymously transmit a message to a receiver of her choosing. The protocol first deals with potential collisions, verifying whether there is a single sender. Then the sender anonymously notifies the receiver only that he/she is about to receive a message, followed by the message transmission. The encoding of the message maps a $m$-bit string onto a $m'=m+2\gamma$-bit string, where $\gamma < \beta+\log(m+1)$. The decoding process reveals whether the message has been tampered with, with success probability $1-2^{-\beta}$ \cite{cramer2008detection}. An explicit construction of the encoding and decoding is given in Appendix.~\ref{sec:message}. 
The protocol follows.
\begin{enumerate}
    \item The participants execute the collision detection protocol, and continue if there is a single sender.
    \item Denote the unique sender as $S$ and the receiver as $R$. The participants perform the \textit{notification} protocol, where $S$ notifies that $R$ is to receive a message. 
    \item The following message transmission protocol is executed:
    \begin{enumerate}
    \item The sender encodes the message using an algebraic manipulation and detection code (see Appendix), which maps the message $M$ (of length $m$ bits) into $M'$, which has length $m' \approx m+2(\log m + \beta)$  bits. 
    \item The participants perform  $m'$ rounds of the anonymous broadcasting protocol, where $S$ uses $M'$ as the input, 
    $R$ uses a random $m'$-bit string $r$ as the input, and all other participants input 0 at every round. 
    \item Let $d$ be the output of the $m'$ anonymous broadcasting protocol. The receiver computes 
    $M''= d \oplus r$.
    \item A veto protocol is performed. Everyone inputs 0 except for the receiver, who inputs 1 if an error is detected in the message, otherwise she inputs 0.
    If the output of veto is 1, this broadcasts to the sender that the message has been corrupted.
\end{enumerate}
    \end{enumerate}

\begin{table}[t]
\begin{center}
\begin{tabular}{ | c |  c|c |   } \hline 
 Protocol &    \makecell{No. of secret bits \\consumed} &   \makecell{Success probability\\ (minimum)}\\
 \hline
 Broadcasting                           & $n(n-1)/2$             &  1\\ \hline
 Veto                             & $\beta n^2(n-1)/2$     & $1-2^{-\beta}$ \\ \hline
 Notification                     & $\beta n^2(n-1)/2$     & $1-2^{-\beta}$  \\  \hline
 \makecell{Collision \\ detection} & $\beta n^2(n-1)  $    &  $\left(1-2^{-\beta} \right)^2$ \\ \hline
\makecell{Message \\ transmission \\($m$-bit)}        & 
\makecell{$m+ 2(\log [m] + \beta)$ \\ $+ 2 \beta n^2(n-1)$} & $1-2^{-\beta}$ \\
 \hline
\end{tabular}
\end{center}
\caption{\label{tab:sum} \textit{The number of secret bits and security parameters of protocols implemented in this paper.} Here $n$ is the number of participants and $\beta$ is mutually agreed upon by all participants to achieve a minimum desired success probability.}
\end{table}

For anonymous private message transmission, we define the efficiency of the encoding as $ m/(m+2\gamma) $, which is the ratio of the message length to the required encoding. We plot $m/(m+2\gamma)$ for $\beta = 16$ in Fig.~\ref{fig:eff}. Evidently, the longer the input message, the more efficient the encoding becomes.
For the experiment, we used two encodings. We map a 512-bit and 1024-bit message to a $554$ and 1068-bit encoding respectively, with $\gamma = 21$ (for the 512-bit message) and 22 (for the 1\,kb message). 

\section{Results}
\label{sec:results}

We successfully implemented all 5 anonymous protocols on our quantum network testbed consisting of 8 simultaneously connected users without trusted nodes. Our fully connected network where every user exchanges secure keys with every other user is crucial to the implementation of these protocols. The superconducting detectors used provide roughly 18.4 hours of continuous operation before the cryostats need to be thermally cycled. During this $\sim$ 18-hour run of the network, we chose to implement the basic anonymous broadcasting protocol throughout, then use this to implement the other 4 protocols one by one.

\begin{center}
\begin{figure*}[t]
\centering
\includegraphics[width=2.0\columnwidth]{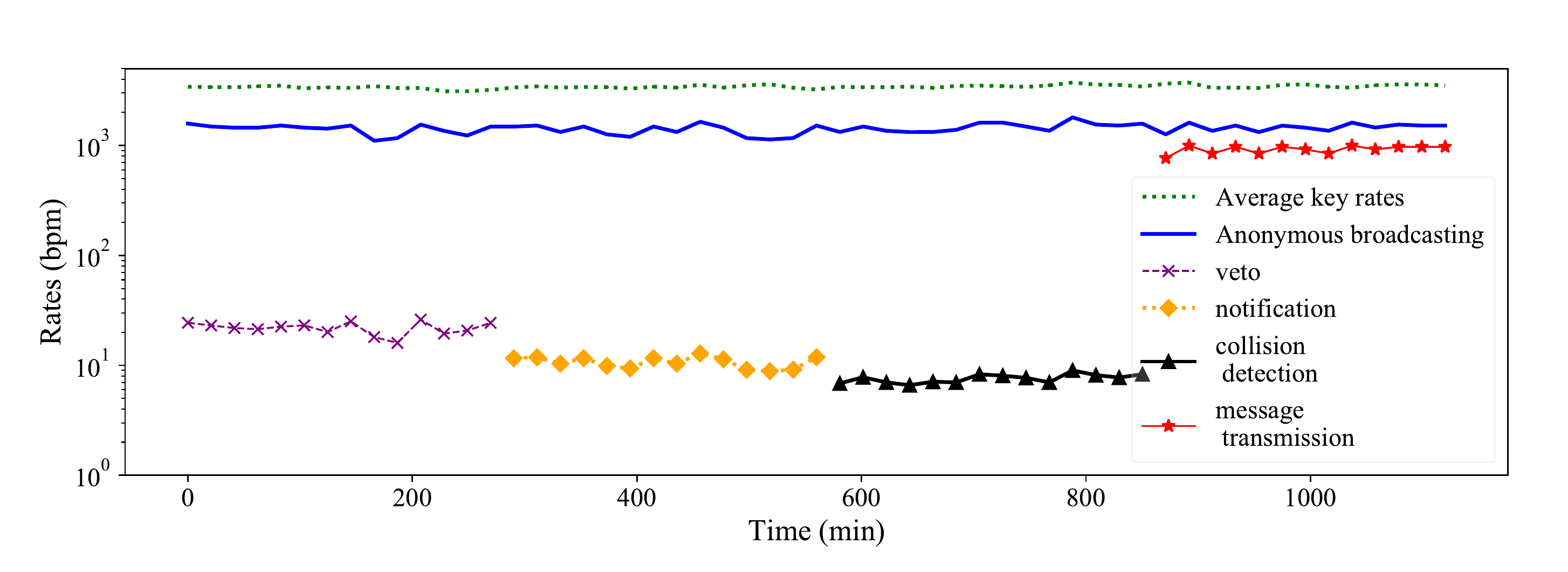}
\caption{\textit{Anonymous communication rates over time for the lab experiment}. We show: the data the average key generation rate across all the links (green dotted line), anonymous broadcasting (parity) (blue solid line) and rate at which veto (purple crosses), notification (yellow diamonds), collision detection (black triangles) and message transmission (red stars) were performed. To ensure a minimum success probability of $>$ 99.99\,\% (1 for veto) we used $\beta = 16$. Here the keys are generated every {20} mins while including finite key effects. 
} \label{fig:rates}    
\end{figure*}
\end{center}

To demonstrate the stability of our network, we show the rate at which the protocols can be performed over the 18.4 hours in Fig.~\ref{fig:rates}. To be able to account for finite key effects with a security parameter of {$10^{-5}$}, we computed the private key once every 20 minutes. For each protocol, we use keys generated over a total time of 280 minutes. 

We implemented the veto protocol, for a random mixture of input 0 and 1's, likewise for the collision detection protocol. The veto protocol consumes the equivalent of only a few rounds of the anonymous broadcasting protocol if the input is 1, and $n\beta$ rounds if the input is 0. The notification protocol consumes $n\beta$ rounds of anonymous broadcasting regardless of the input. We executed the anonymous private message protocol using a mixture of 1~kb and 0.5~kb message lengths.

Due to the nature of the anonymous broadcasting protocol, the rate at which the protocol can be implemented is limited by the slowest link. This can be seen in the disparity between the average key rate and the anonymous broadcasting rate.

 For anonymous private message transmission, the protocol requires running an anonymous collision detection, veto and notification step. Thus the total number of bits consumed is larger. 
The communication rate is $m/(m+ 2\log m + 2\beta + 2 \beta n^2(n-1) )$, where the term of $2 \beta n^2(n-1)$ are the overheads associated with the collision detection, notification and veto protocols required as part of the protocol. This rate can be made arbitrarily close to 1 by choosing $m$ large enough.

A summary of the number of secret bits consumed by each protocol that we implemented is given in Table~\ref{tab:sum}. We note that our protocols are tolerant to errors in the secure keys and we discuss how in Appendix~\ref{sec:err}.

\section{Conclusions}

Quantum communication networks have largely focused on key distribution. Here we have shown how these networks can be used to protect user privacy, a task often just as important as ensuring that a transmitted message is secure. 
In this work, we have implemented a set of 5 information-theoretically secure anonymous protocols -- broadcasting, veto, notification, collision detection and anonymous private message transmission on an 8-user quantum network. These protocols can all be used as primitives for a wide range of applications.


The anonymous broadcasting protocol has a throughput equal to the minimum key rate of the quantum network and serves as a basis for most other anonymous protocols. The protocol requires almost no classical communication overhead to implement except for announcing the final parity.
The veto protocol consumes only a few rounds of anonymous broadcasting if the input is 1; it requires $n\beta$ rounds if the input is 0. The notification protocols also consume $n\beta$ broadcasting runs.
We believe that these protocols will prove to be useful for voting based applications and truly private social media.
Collision detection must become an important part of any anonymous network application and we believe that this could form the basis of a completely anonymous network control plane which could optimally allocate resources and communication bandwidth in a network without collecting individual users' network usage information.
  
We anticipate that the most useful protocol would be anonymous private message transmission, which allows one user to send an arbitrary-sized message to another user in secret. This has important implications for privacy-preserving applications, such as tipping off the police anonymously \cite{PhysRevA.97.032345}, secret voting, secure electronic auctions \cite{stajano1999cocaine}, anonymous cryptocurrency transactions \cite{ruffing2017p2p}, multi-party computation \cite{movahedi2014secure} etc. Here we provide an explicit construction for the encoding and decoding of the message.

The above protocols only preserve anonymity within a single quantum network and their usefulness will grow as the size of the network increases. Maintaining anonymity in networks where active switching is used to choose which users can communicate at any given point in time or in scenarios where independent quantum networks are interconnected remains an open challenge.

We have used secret keys generated via QKD to perform protocols that allow for complete anonymity in a communication network.
Our experiment represents one of the first quantum communications protocols beyond point-to-point QKD on a quantum network. Our experiment thus demonstrates the capabilities of a quantum network to realise non-trivial security tasks even when quantum memories are not available.

\begin{acknowledgments}
The research leading to this work has received funding from the Engineering and Physical Science Research Council (EPSRC) Quantum Communications Hubs EP/M013472/1 \& EP/T001011/1 and equipment procured by the QuPIC project EP/N015126/1. We acknowledge the Ministry of Science and Education (MSE) of Croatia, contract No. KK.01.1.1.01.0001.
We acknowledge financial support from the Austrian Research Promotion Agency (FFG) project ASAP12-85 and project SatNetQ 854022.
This work was partially supported by the European Union’s Horizon 2020 research and innovation programme under the Marie Sklodowska-Curie grant agreement number 675662 (QCALL).
The authors would also like to thank Peter P. Rohde for insightful discussions.
\end{acknowledgments}
\begin{center}
 \section*{Author Contributions}   
\end{center}
The theoretical aspects of this work was led by ZH with help from CL and AOQ. The experimental team was led by SKJ with DA, SW, ML, SPN, ZS, LK, MS, RU and JGR. The software for processing the keys and implementing the algorithms was written by NV, ZH and BL with help from SKJ. The user modules were built by ML, ZS and MS, the source by SW, SPN, SKJ and RU. The multiplexing and distribution scheme was built by DA. Funding for this work was obtained by JGR with help from SKJ and DA. SKJ coordinated the project. The paper was written by ZH, SKJ, DA, CL and AOQ, the other authors proofread. All authors discussed the results and commented on the manuscript. ZH and SKJ contributed equally to this work and are joint first author.

\bibliography{qab} 

\appendix

\section{Notification}\label{sec:notificaion}

The notification protocol allows for a participant to {notify} a list of others, but revealing no information on neither the number of participants being notified nor the identity of the sender.

For $n$ participants, each participant $j$ has a $n$-bitstring  $(x^1_j, \dots, x^n_j)$. If participant $j$ wants to notify $i$, he sets $x^{i}_j=1$, and $0$ otherwise. 
For each 
participant $i$ waiting to be notified, the following is repeated $\beta$ times

\begin{enumerate}
    \item Each participant $j \neq i$ sets the value of another bit, $c_j^i$, in the following way: if $x_j^i = 0$ , then $c_j^i$ = 0; 
    if $x_j^i = 1$, then $c_j^i$ is chosen randomly.
    \item The anonymous broadcasting protocol is executed with inputs $c_1^i,c_2^i...c_n^i$, with the exception that participant $i$ does not broadcast.
    \item  Participant $i$ computes the overall parity in secret; he is \textit{notified} if the parity is 1 for any run of the protocol.
\end{enumerate}
Each round of the notification protocol consumes at most $\beta n^2(n-1)/2$ secret bits, and succeeds with probability at least $1-2^{-\beta}$.

\section{Collision detection} \label{sec:collision}
The collision detection protocol enables the participants to verify whether or not there is a single sender in the group. This can be used as a procedure for the implementation of anonymous private message transmission. 
The protocol is divided into two sub-routines A and B.

\noindent
Step \textbf{A}:
\begin{enumerate}
    \item  All participants perform the veto protocol. If they (do not) want to transmit a message, they choose input $x_i =1 (0)$.
    \item Everyone takes note of the logical OR of the other participants’ inputs.
    \item If the outcome of the above is 0, no one wants to transmit a message, and we terminate.
\end{enumerate}

\noindent 
Step \textbf{B}:
\begin{enumerate}
    \item Each participant $i$ chooses the value of another bit, $b_i$ in the following way: from \textbf{A}, 
    if he had detected that another participant had input 1, 
    he chooses $b_i = 1$; otherwise he sets $b_i = 0$.
    \item All participants perform another veto protocol with inputs $b_i$. That is, he uses $b_i$ to choose the value of $c_i$ to feed into the veto protocol.
    \end{enumerate}
The three outcomes for the collision detection protocol are:
\begin{equation}
    \begin{cases}
      0, & \text{if the output of  \textbf{A} is } 0 \\
      1, & \text{if the output of  \textbf{A} is 1 and the output of \textbf{B} is 0} \\
      2, & \text{if the output of  \textbf{A} is 1 and the output of \textbf{B} is 1}. \\
    \end{cases}
  \end{equation}
\noindent These correspond to no sender, single sender, and multiple senders respectively.
The collision detection protocol consumes at most $\beta n^2(n-1)$ secret bits, and succeeds with probability at least  $(1-2^{-\beta})^2$.

\section{Anonymous private message transmission}
\label{sec:message}
The anonymous message transmission allows a sender to anonymously transmit a message to a receiver of his choosing. It first deals with potential collisions, verifying whether there is a single sender. Then the sender notifies the receiver that he is about to receive a message, followed by the message transmission:
\begin{enumerate}
    \item The participants execute the collision detection protocol. They continue if the outcome is 1 (single sender), otherwise they output 0 or 2 (no transmission or collision);
    \item Denote the unique sender as $S$, and the receiver as $R$. They perform the notification protocol, where $S$ inputs the bit $x_S^R=1$, and 0 otherwise. All other participants input $x_i^j = 0$;
    \item The \textbf{fixed role anonymous message transmission} protocol is then executed.
\end{enumerate}

\vspace{3mm}
\subsection{Fixed role anonymous message transmission}
In the fixed role anonymous message transmission protocol, the sender $S$ and the receiver $R$ are aware of their own roles.
The task is for the sender to correctly transmit a $m$-bit long message, $M$.
\begin{enumerate}
    \item The sender encodes the message using an algebraic manipulation and detection (AMD) code, which encodes the message $M$ (of length $m$ bits) into $M'$ (of length $m'= m+2\gamma$ ) bits. (See the following subsection.)
    \item The participants perform  $m'$ rounds of the anonymous broadcasting protocol, where $S$ uses $M'$ as the input, $R$ uses a random $m'$-bit string, r, and all other participants input 0. 
    \item Let $d$ be the output of the $m'$ anonymous broadcasting protocols. The receiver computes 
    $M''= d \oplus r$.
    \item A veto protocol is performed. Everyone inputs 0 except for the receiver, who inputs 1 if an error is detected in the message, otherwise she inputs 0.
    If the output of veto is 1, this broadcasts to the sender that the message has been corrupted and the protocol is aborted.
\end{enumerate}

\onecolumngrid

\subsection{Algebraic manipulation and detection (AMD) code}

The sender $S$ wants to send the message $M$ to the receiver $R$. We require that $R$ is able to assess whether the message he receives has been corrupted. In order to do so, $S$ encodes the message $M$ and sends to $R$ the encoded message $M'=\mathcal{F}(M)$. When $R$ receives what passes through the channel, $N=\mathcal{E}{(M')}$, he decodes it using the function $\mathcal{G}(N)$. The function $\mathcal{G}(N)$ announces that there is an error if the message has been corrupted during transmission and the original message $M$ otherwise. The function $\mathcal {G}$ fails (and therefore the protocol fails) with probability $\le 2^{-\beta}$, were $\beta$ is the security parameter.

We now give an explicit way of constructing $M'$ for a binary bitstring of length m bits, given $M$ and the parameter $\beta$. $M'$ consists of 3 blocks, the first block is $M$, the second block, $\Gamma$, is a random bitstring of length $\gamma$ (defined below), and the third block is another bitstring F of length $\gamma$

\begin{align*}
M'= (\underbrace{1, 1, 0, 0, 1, 1, 0, 0, 1, 1,...}_M,
     \,\,
     \underbrace{0, 1, 0, 0,0,0,0}_\text{$\theta$, $\gamma$-bit random string}, \,\,
     \underbrace{0, 1, 1, 0, 0, 1, 0, 0}_\text{$\tau=F(M,\theta)$, $\gamma$ bits}).
\end{align*}

\begin{enumerate}
    \item Find $d$, the smallest odd integer such that:
        \begin{align} \label{dm}
        d(\beta+ \log_2(d+1))\ge m
        \end{align}
    Define  $\gamma = \lceil \beta + \log(d+1)\rceil$, where $\lceil \cdot \rceil$ is the ceiling function.
    \item Find an irreducible binary polynomial $b(x)\in \mathbb F_2[x]$ of degree $\gamma$. 
    \item Generate a random bitstring, $\theta$, of length $\gamma$.
    \item Compute $F(M,\theta)$ as explained in the next section.
    \item Concatenate $\theta$ and $F(M,\theta)$ to $M$, this is now $M'$.
\end{enumerate}

\subsection{How to compute the function $F$}
The function $F$ takes as input a bit string $(\mu_1, \dots, \mu_m, \theta_1, \dots, \theta_\gamma)$ of length $m + \gamma$ and outputs a bit string $(\tau_1, \dots, \tau_\gamma)$ of length $\gamma$. The function $F$ uses finite field arithmetic, we assume that we have access to the global parameters of the system (i.e. the irreducible binary polynomial $b(x)$) when we have to compute $F$. The explicit construction follows.
\begin{enumerate}
\item Write the string $\mu = (\mu_1, \dots, \mu_m)$ as a string of $d$ chunks each of length $\gamma$ padding with $d \gamma - m$ zeros if necessary. Note that by Eq.~\ref{dm} $d\gamma \ge m$. In other words, we write $\mu$ as 
\begin{align*}
(\mu_1, \dots, \mu_{\gamma} | \mu_{\gamma+1} \dots \mu_{2\gamma}| \dots |\mu_{(d-1)\gamma + 1}\dots \mu_{m}, 0 \dots 0)
\end{align*}
and we refer to any of the $d$ chunks of length $\gamma$ as $\mu^{[0]}, \dots, \mu^{[d-1]}$. Any of these chunks $\mu^{[i]}$ can be seen as an element of the field $\mathbb F_{2^\gamma}$ by the polynomial representation in $\mathbb F_2 [x]$:
\begin{align}
\label{eq:mu}
\mu^{[i]}=(\mu_{i\gamma+1}, \dots, \mu_{(i+1)\gamma}) \rightarrow \mu_{i\gamma +1} x^{\gamma-1}+ \mu_{i\gamma +2}x^{\gamma-2} + \dots + \mu_{(i+1)\gamma}. 
\end{align}
In the following, with slight abuse of notation, we will indicate with $\mu$ both the bitstring and its polynomial representation.
\item Write the string $(\theta_1, \dots, \theta_{\gamma})$ as a binary polynomial of degree at most $\gamma-1$ as done for $\mu$ in Eq.~\ref{eq:mu}:
\begin{align}
\theta(x) = \theta_1 x^{\gamma-1} + \theta_2 x^{\gamma-2} + \dots + \theta_{\gamma}
\end{align}
\item Again using arithmetic modulo $2$ (i.e. XOR) compute the binary polynomial $f(x)$:
\begin{align*}
f(x) &= f(\mu, \theta)\\
&=f(\mu^{[0]}(x), \dots, \mu^{[d-1]}(x), \theta(x))\\
&= \theta(x)^{d+2} + \sum_{i = 1}^d \mu^{[i-1]} (x)\theta(x)^i.
\end{align*} 
\item Reduce the polynomial $f(x)$ modulo $b(x)$, i.e. compute the remainder $r(x)$ of $f(x)/b(x)$ such that
\begin{align*}
f(x) = b(x)h(x) + r(x)
\end{align*}
and the degree of $r(x)$ is strictly less that the degree $\gamma$ of $b(x)$
\item Find the binary vector $\tau$ of length $\gamma$ corresponding to the polynomial $r(x)= r_1 x^{\gamma-1} + r_2 x^{\gamma-2} + \dots, r_{\gamma}$
\begin{align*}
(\tau_1, \dots, \tau_{\gamma}) =( r_1, \dots, r_{\gamma})
\end{align*}
\item Output $(\tau_1, \dots, \tau_{\gamma})$
\end{enumerate}

\subsection{Decoding operation by the Receiver}

The receiver obtains the bit string $\rho = (\tilde \mu_1, \dots, \tilde \mu_m, \tilde \theta_1, \dots, \tilde \theta_\gamma, \tilde \tau_1, \dots, \tilde \tau_\gamma)$ of length $m + 2\gamma$. He computes $\mathcal G(\rho)$:
\begin{enumerate}
\item He computes the function $F(\tilde \mu, \tilde \theta_1, \dots, \tilde \theta_m)= (\iota_1, \dots, \iota_\gamma)$ exactly the same way the sender did.
\item If $(\iota_1, \dots, \iota_\gamma) = (\tilde \tau_1, \dots, \tilde \tau_{\gamma})$ he retrieves the message $\tilde \mu_1, \dots, \tilde\mu_m$ and he knows that this is equal to the message $(\mu_1, \dots, \mu_m)$ that the sender intended for him to receive with high probability (i.e. $\ge 1- 2^\beta$). In this case the output is $(\tilde\mu_1, \dots, \tilde\mu_m)$.
\item If not, he flags that the message has been corrupted, and the output is $\bot$, i.e. ``there was an error''.
\end{enumerate}


\subsection{Encoding a 1-kilobyte message}

In the actual protocol, we encoded a 1kb (1024 bit) message with parameter $\beta = 16$. We have the following global parameters:

\begin{align}
 &d ( 16 + \log_2(d+1)) \geq 1024 \nn
\Rightarrow d &= 49.\nn
\gamma &= \ceil{ 16+\log_2(50)}  = 22,\nn
b(x) &= x^{22} + x +1
\end{align}


%
%
 %
%

%
\section{Tolerating errors in the QKD keys} \label{sec:err}


The anonymous protocols assume that the private keys shared between all the users are errorless. However, this may not be the case for keys generated via QKD in general. It may even be advantageous to reduce the amount of error correction performed to allow a small probability of error in the final key to increase the overall key generation rate when using the keys for the anonymous protocols presented here. The security concerns when doing so will need careful examination but we believe it should be possible. Nevertheless, our protocols can be made robust against such errors.
Given $n$ participants (nodes), there are $n(n-1)/2$ links, representing the number of shared secret keys.
In each round of the anonymous broadcasting protocol, an error in this protocol will be caused by having an odd number of errors in the $n(n-1)/2$ links.  

 Assume for simplicity that all the shared keys have error probability
$r_e$. Then, the probability of error for each run of the anonymous broadcasting protocol, $E_\text{parity}$, is given by
\begin{align}
E_\text{parity} &= \sum_{j \text{ odd}} ^ {M} {M \choose j} ~r_e^j~ (1- r_e)^{M-j}, \quad M= n(n-1)/2 \nn
             &= \frac{1}{2} \left[1-(1-2 r_e)^{\frac{1}{2} n(n-1) }\right].
\end{align}
\noindent where $n$ is the number of participants.

When fed into the \text{veto} protocol, having an error will affect the two {veto} outcomes differently: the veto protocol is designed such that if any run of the protocol is equal to 1, then the output is set to 1. The error will leave a veto with input 1 essentially unchanged, since the probability that it will flip all the 1's in the protocol is negligible.
However, if the input to the veto protocol is 0, then the probability of no error occurring becomes $(1-E_\text{parity})^{n \beta}$, which is problematic when $n$ becomes large.
Let us consider an error rate which is in practice very large, say an error rate of {$10^{-4}$}. This corresponds to a parity error rate of $3\times 10^{-3}$. One can, for example, reduce $r_e$ by using a repetition code. This is done by performing each round of the anonymous broadcasting protocol with the exact same inputs $N$ times, which comes at the expense of reducing the broadcasting rate by a factor of $N$. Since the correction capability of a length $N$ repetition code is $d = \lfloor (N-1)/2 \rfloor$, the parity error rate is then reduced to: 
\begin{align}
\label{eq:repetition}
 E'= &\sum_{i=d+1}^N {N \choose i }E_\text{parity}^i (1-E_\text{parity})^{N-i}.
\end{align}
For $N=5$ and $d=2$ Eq.~\ref{eq:repetition} yields $E' \approx 2\times 10^{-7}$.
%
%
%
With this correction, the veto protocol can have a very high success probability even for many user networks. For example when $n=8$, the veto protocol succeeds with probability 
$(1-E')^{8\beta} \approx 0.99997$, with $\beta = 16$.

\end{document}